\documentclass[%
 aip,
 amsmath,amssymb,
 reprint,%
]{revtex4-1}

\usepackage{graphicx}
\usepackage{dcolumn}
\usepackage{bm}

\usepackage[utf8]{inputenc}
\usepackage[T1]{fontenc}
\usepackage{mathptmx}

\begin{document}

\preprint{AIP/123-QED}

\title{Experimental implementation of bias-free quantum random number generator based on vacuum fluctuation}

\author{Ziyong Zheng}

\author{Yichen Zhang}%

\author{Song Yu}
\email{yusong@bupt.edu.cn.}
\affiliation{State Key Laboratory of Information Photonics and Optical Communications, Beijing University of Posts and Telecommunications, Beijing, 100876, China}

\author{Hong Guo}
\affiliation{State Key Laboratory of Advanced Optical Communication Systems and Networks, School of Electronics Engineering and Computer Science, and Center for Quantum Information Technology, Peking University, Beijing, 100871, China}%

\date{\today}

\begin{abstract}
We experimentally demonstrate a bias-free optical quantum random number generator with real-time randomness extraction to directly output uniform distributed random numbers by measuring the vacuum fluctuation of quantum state. A phase modulator is utilized in the scheme to effectively reduce the influence of deviations between two arms of the generator caused by the imperfect practical devices, which is an innovative solution in the field of quantum random number generator. In the case where the feedback modulation frequency is much faster than the phase jitter, an unbiased result can be obtained by an additional subtraction between the compensation signal and its average value to eliminate residual deviation. A following randomness extractor is applied to eliminate the influence of residual side information introduced by the imperfect devices in practical system.
\end{abstract}

\maketitle

	\section{\uppercase{Introduction}}

	\noindent Random numbers are widely used in simulation \cite{Ferrenberg1992}, lottery, cryptography \cite{Gennaro1621063} and other applications. The randomness of random numbers has a significant impact on the performance of the whole system. Especially in cryptography applications, random numbers with poor randomness will directly reduce the security of the cryptography system \cite{Bouda2012Weak}. The rapid development of quantum cryptography technologies such as quantum key distribution  \cite{Weedbrook2012Gaussian,Scarani2012The,Diamanti2016Practical,Zhang2017Continuous,Gisin2002quantum} which require secure random number generation, unarguably accelerate the researches about true random number generation. Quantum random number generator (QRNG) exploits intrinsic probabilistic quantum processes to directly generate true random numbers, which is regarded as a promising technology  \cite{Ma2016Quantum,Bera2017Randomness,Herrero2017quantum,Jennewein2000A}. Therefore, many related works have been put forward in recent years. These schemes use quantum sources includes photon path  \cite{Jennewein2000A,Andr2000Optical}, photon arrival time  \cite{Michael2009Photon,Nie2014Practical,Dynes2008A,Wahl2011An,Ma2005Random}, photon number  distribution\cite{Wei2009Bias,Furst2010High,Applegate2015Efficient,Ren2011Quantum}, vacuum fluctuation \cite{Gabriel2010A,Shen2010Practical,Symul2011Real,Haw2015Maximization,Zhou2017Practical,Santamato2017An,Bingjie2017High,ZHENG20186Gbps}, phase noise \cite{Qi2010Highspeed,Guo2010Truly,Xu2012Ultrafast,Abellan2014Ultra,Nie2015The,Yang2016A,Zhang2016Note,Liu2017117} and amplified spontaneous emission noise of quantum states \cite{Williams2010Fast,Li2011Scalable,Martin2015Quantum,Liu2013Implementation,Wei2012High}, etc. Typically, protocols based on the measurement of vacuum fluctuation are more applied and valuable QRNG protocols, for its convenience of state preparation, insensitivity of detection efficiency and high generation speed.

	The first QRNG based on vacuum fluctuation is proposed in 2010 by measuring the quadrature of the vacuum state, which can be expressed as $\left| 0 \right\rangle {\rm{ = }}\int_{{\rm{ - }}\infty }^\infty  {\psi \left( x \right)} \left| x \right\rangle dx$ in the quadrature representation, where $\left| x \right\rangle $ is the amplitude quadrature eigenstates and ${\psi \left( x \right)}$ is the ground-state wavefunction, which is a Gaussian function centered around $x = 0$ \cite{Gabriel2010A}. 
	
	Ideally, the electrical signal output from the detector should be evenly distributed near the 0 value. While in practical system, the deviation of two signals output from the homodyne detector caused by imperfect unbalanced devices, such as asymmetric beam splitter or photodiodes (PDs) with different response efficiency, will often cause the saturation of detector, which is a great challenge for practical system. Generally, symmetrical devices will be chosen as symmetrical as possible so as to reduce the deviation of the two arms. However, the practical devices can not achieve complete symmetry. Minor deviation will be amplified to a large voltage by the homodyne detector with a large gain, which will lead to saturation of homodyne detector.

	Protocol introduced in Ref. \cite{Shen2010Practical,Symul2011Real,Haw2015Maximization} utilized frequency shift and filtering technology to obtain the signal in the required frequency band and filter out the signal outside the band, including low frequency signals that causes signal deviation. While it is based on the premise of unsaturated signal. When the detector is saturated, in fact, this operation can not eliminate the effect of saturation basically caused by front-end parts. An intuitive solution is to introduce an adjustable attenuator at both output ends of the beam splitter so as to adjust the balance of the two arms. However, the mechanical jitter of the practical attenuator will inevitably lead to imbalance and be greatly amplified by the amplifier in the detector, so that the detector will still be saturated. Protocol introduced in Ref. \cite{Bingjie2017High,ZHENG20186Gbps} proposed an improved solution that is using $AC$ coupling detector to suppress the low-frequency components of the signals detected by the PDs. In this way, the $DC$ components of the difference between the electrical signals output from the two PDs will firstly be filtered out and then the signals in the remaining band will be amplified. To some extent, the feasibility of this scheme depends on the perfect filtering of low frequency components by transimpedance bandpass amplifier. However, in practice, the imperfection of the filter can not eliminate the influence of low frequency signals perfectly, which causes the signal still to be affected by residual low-frequency jitter. Therefore, how to achieve an effective and feasible deviation elimination method is a meaningful and practical problem.
	
	We experimentally demonstrate a bias-free optical quantum random number generator with real-time randomness extraction to directly output uniform distributed random numbers by measuring the vacuum fluctuation of quantum state. The generator utilizes a phase modulator to effectively reduce the deviation between two arms of the generator caused by the imperfect practical devices. Unbiased results can be obtained by subtracting the mean value from the compensated signal in every modulation period since the feedback modulating frequency is far faster than the phase jitter. Thus our generator can output bias-free and real-time random numbers stably at a speed of 640 Mbps by applying a real-time randomness extractor to eliminate the influence of classical noise.

\begin{figure*}[t]
		\centering
		\includegraphics[width = 14 cm]{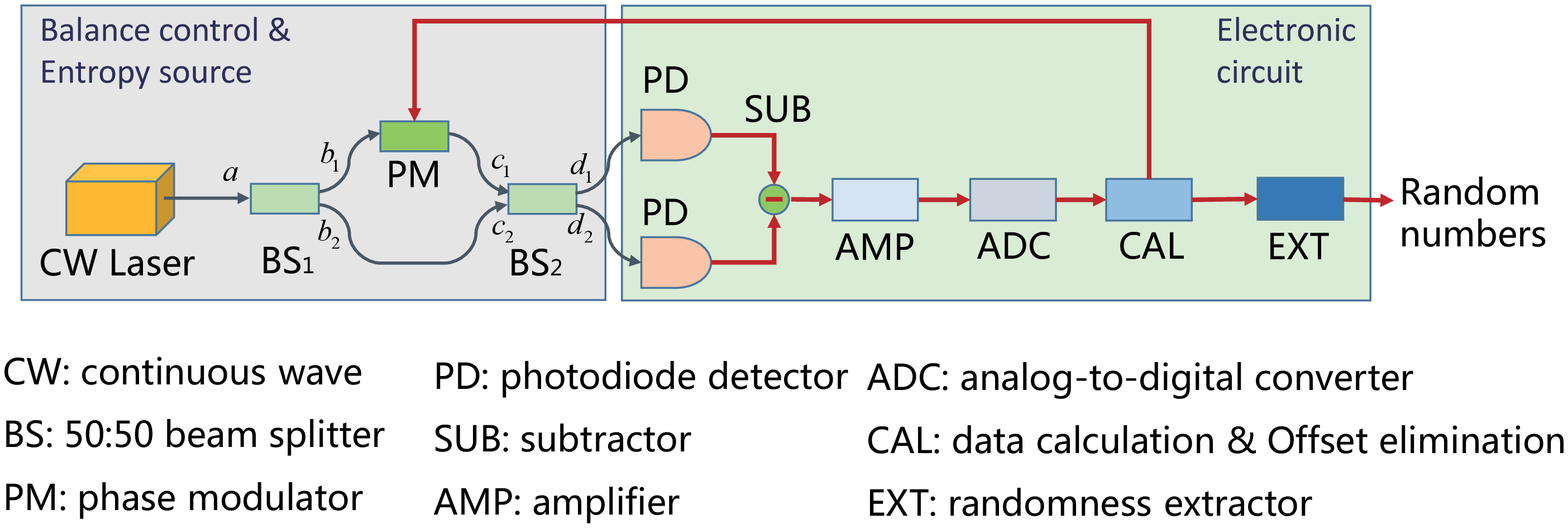}
		\caption{\label{system_scheme}Scheme of the bias-free QRNG based on vacuum fluctuation. The $CW$ beams emitted by the laser diode is divided into two beams by the first balanced beam splitter and one arm is modulated by a phase modulator to maintain the phase difference between the two arms as a fixed value. The interference result of the second beam splitter would be a stable value.}
	\end{figure*}

	\noindent The QRNG proposed in Ref. \cite{Gabriel2010A} essentially exploits the quantum uncertainty of continuous observables, which is quadrature amplitude of vacuum state to generate true random numbers. The measurement of the quadrature amplitude collapses the ground-state wave function, which is a Gaussian function centered around $x = 0$, into quadrature eigenstate. While the practical imperfect devices will make the output of the two PDs different, so that there will be a deviation after the subtraction of the two electrical currents. To eliminate this deviation, a scheme based on phase modulation is proposed with reference to laser interferometry technology. The block diagram of the scheme is shown in Figure. \ref{system_scheme}. 
	
	The first beam splitter ${(BS_1)}$ with three ports divides the light beam from the $CW$ laser into the upper and lower arms with a transmission coefficient of ${\eta _{ab_1}}$ and ${\eta _{ab_2}}$. A phase modulator ${(PM)}$ with insertion loss of ${\eta _{PM}}$ is connected to the upper arm. The two output signals are then connected to the input ports of the second beam splitter ${(BS_2)}$. Four parameters named ${\eta _{c_1d_1}}$, ${\eta _{c_1d_2}}$, ${\eta _{c_2d_1}}$, ${\eta _{c_2d_2}}$ are used to represent the transmission coefficients of port ${c_1}$ to ${d_1}$, ${c_1}$ to ${d_2}$, ${c_2}$ to ${d_1}$ and ${c_2}$ to ${d_2}$. The efficiency of the photoelectric conversion of the two PDs is labeled ${\eta _{pd_1}}$ and ${\eta _{pd_2}}$ respectively.
	
	When the vacuum noise is not considered, from a classical point of view, it is intuitive that there is a phase difference $\Delta \varphi $ between the upper and lower arms. The output photocurrent of $PD_1$ can be expressed as 
	
	\begin{equation}
	\label{eq1}
	\begin{split}
	{i_{p{d_1}}}{\rm{ }} &= |\sqrt {{\eta _{p{d_1}}}} ({\eta _{a{b_1}}}{\eta _{{c_1}{d_1}}}{\eta _{pm}}{E_{in}}{e^{j\Delta \phi }} + {\eta _{a{b_2}}}{\eta _{{c_2}{d_1}}}{E_{in}}){|^2} \\&= {\eta _{p{d_1}}}{E^2}_{in}({\eta ^2}_{a{b_1}}{\eta ^2}_{{c_1}{d_1}}{\eta ^2}_{pm} + {\eta ^2}_{a{b_2}}{\eta ^2}_{{c_2}{d_1}}) \\&+ 2{\eta _{p{d_1}}}{E^2}_{in}{\eta _{a{b_1}}}{\eta _{{c_1}{d_1}}}{\eta _{pm}}{\eta _{a{b_2}}}{\eta _{{c_2}{d_1}}}\cos (\Delta \phi ).
	\end{split}
	\end{equation}
	
	Similarly, the photocurrent output from PD2 can be expressed as
	
	\begin{figure*}[t]
		\centering
		\includegraphics[width = 14 cm]{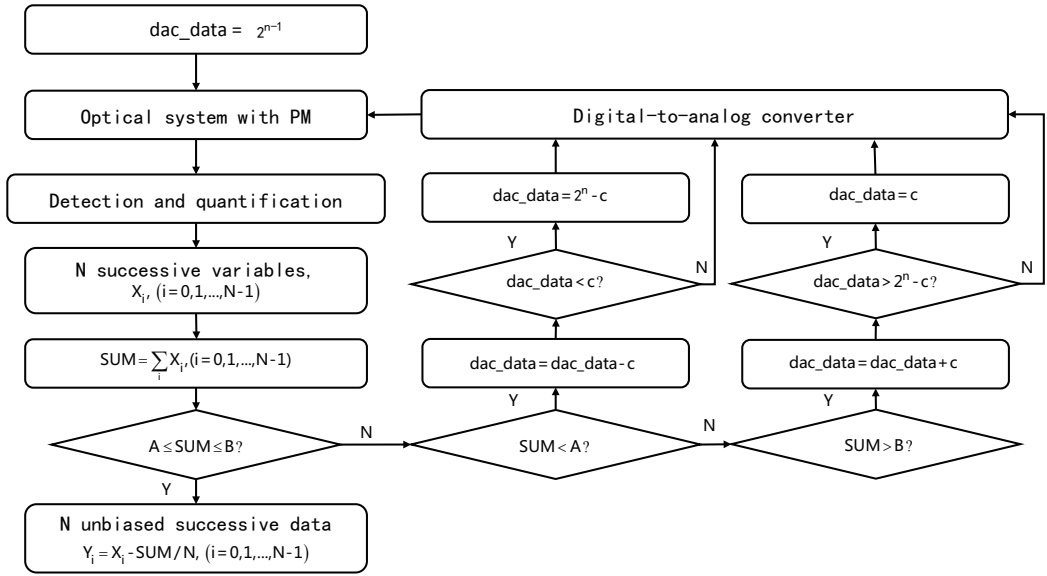}
		\caption{\label{algorithm_pm}Algorithm of deviation elimination progress. We initialize the value of $dac\_data$, which represents the digital data used to drive the $DAC$, to 8092. By comparing the sum of the samples collected in a time interval $\tau $ with the preset interval, the phase difference between the two arms of the system will calculated and a feedback compensation operation will be performed with a compensation frequency far greater than the phase jitter to make the fluctuation of the interference results be stabilized in a very small interval. And $c$ is the step of adjusting $dac\_data$.}
	\end{figure*}
	
	\begin{equation}
	\label{eq2}
	\begin{split}
	{i_{p{d_2}}} &= |\sqrt {{\eta _{p{d_2}}}}({\eta _{a{b_1}}}{\eta _{{c_1}{d_2}}}{\eta _{pm}}{E_{in}}{e^{j\Delta \phi }} + {\eta _{a{b_2}}}{\eta _{{c_2}{d_2}}}{E_{in}}){|^2}\\&
	= \eta _{p{d_2}}{E^2}_{in}({\eta ^2}_{a{b_1}}{\eta ^2}_{{c_1}{d_2}}{\eta ^2}_{pm} + {\eta ^2}_{a{b_2}}{\eta ^2}_{{c_2}{d_2}})\\&
	+ 2{\eta }_{p{d_2}}{E^2}_{in}{\eta _{a{b_1}}}{\eta _{{c_1}{d_2}}}{\eta _{pm}}{\eta _{a{b_2}}}{\eta _{{c_2}{d_2}}}\cos (\Delta \phi ).
	\end{split}
	\end{equation}

	So the actual current obtained by the homodyne detector will be

	\begin{equation}
	\label{eq3}
	\begin{split}
	i &= {i_{p{d_1}}} - {i_{p{d_2}}}\\&
	{\rm{ = }}{\eta }_{p{d_1}}{E^2}_{in}({\eta ^2}_{a{b_1}}{\eta ^2}_{{c_1}{d_1}}{\eta ^2}_{pm} + {\eta ^2}_{a{b_2}}{\eta ^2}_{{c_2}{d_1}}) \\&- {\eta }_{p{d_2}}{E^2}_{in}({\eta ^2}_{a{b_1}}{\eta ^2}_{{c_1}{d_2}}{\eta ^2}_{pm} + {\eta ^2}_{a{b_2}}{\eta ^2}_{{c_2}{d_2}})\\&+2{\eta _{pm}}{E^2}_{in}\cos (\Delta \phi )({\eta }_{p{d_1}}{\eta _{a{b_1}}}{\eta _{{c_1}{d_1}}}{\eta _{a{b_2}}}{\eta _{{c_2}{d_1}}} \\&- {\eta }_{p{d_2}}{\eta _{a{b_1}}}{\eta _{{c_1}{d_2}}}{\eta _{a{b_2}}}{\eta _{{c_2}{d_2}}}),
	\end{split}
	\end{equation}
	which indicates that $i$ is closely related to the parameters of the devices in the system. To obtain a bias-free $i$, an intuitive solution is to make 
	
	\begin{equation}
	\label{eq3}
	\begin{split}
	&{\eta }_{p{d_1}}{E^2}_{in}({\eta ^2}_{a{b_1}}{\eta ^2}_{{c_1}{d_1}}{\eta ^2}_{pm} + {\eta ^2}_{a{b_2}}{\eta ^2}_{{c_2}{d_1}}) -\\& {\eta }_{p{d_2}}{E^2}_{in}({\eta ^2}_{a{b_1}}{\eta ^2}_{{c_1}{d_2}}{\eta ^2}_{pm} + {\eta ^2}_{a{b_2}}{\eta ^2}_{{c_2}{d_2}}) = 0
	\end{split}
	\end{equation}
	and
	\begin{equation}
	\label{eq4}
	\begin{split}
	&{\eta }_{p{d_1}}{\eta _{a{b_1}}}{\eta _{{c_1}{d_1}}}{\eta _{a{b_2}}}{\eta _{{c_2}{d_1}}} -\\& {\eta }_{p{d_2}}{\eta _{a{b_1}}}{\eta _{{c_1}{d_2}}}{\eta _{a{b_2}}}{\eta _{{c_2}{d_2}}} = 0
	\end{split}
	\end{equation}
	simultaneously, which is not an easy solution to perfectly achieve in practical systems. A feasible alternative is to control the phase difference $\Delta \varphi$ between two paths satisfies 
	\begin{equation}
	\label{eq5}
	\begin{split}
	\cos (\Delta \phi ) &= [{\eta }_{p{d_1}}{E^2}_{in}({\eta ^2}_{a{b_1}}{\eta ^2}_{{c_1}{d_1}}{\eta ^2}_{pm} + {\eta ^2}_{a{b_2}}{\eta ^2}_{{c_2}{d_1}})\\&
	- {\eta }_{p{d_2}}{E^2}_{in}({\eta ^2}_{a{b_1}}{\eta ^2}_{{c_1}{d_2}}{\eta ^2}_{pm} + {\eta ^2}_{a{b_2}}{\eta ^2}_{{c_2}{d_2}})]/\\&
	[2{\eta _{pm}}{E^2}_{in}({\eta }_{p{d_1}}{\eta _{a{b_1}}}{\eta _{{c_1}{d_1}}}{\eta _{a{b_2}}}{\eta _{{c_2}{d_1}}}\\&
	- {\eta }_{p{d_2}}{\eta _{a{b_1}}}{\eta _{{c_1}{d_2}}}{\eta _{a{b_2}}}{\eta _{{c_2}{d_2}}})].
	\end{split}
	\end{equation}
	
	In this way, the deviation can be effectively suppressed, which can directly allow the increase of the input local oscillation power. This will help to improve the problem that the quantized bits of the following analog-to-digital converter ($ADC$) are wasted caused by signals with limited amplitude.
	
	Compared with the interference phenomenon between the classical strong light beams introduced above, the interference between $LO$ and vacuum state will be different. The vacuum state is symmetrical in the phase space, so the interference output of vacuum state and $LO$ with different phases will remain stable. In practice, the two input ports of the ${BS_2}$ are connected to two light beams. Each light beams will interfere with the other beam together with the vacuum fluctuation introduced by the other port. Suppose the vacuum fluctuation obeys the Gaussian distribution $N(0,\sigma _{vac}^2)$, which means its mean value is 0 and its variance is $\sigma _{vac}^2$. So the interference result of vacuum state from port ${c_2}$ and $LO_1$ from port ${c_1}$ will follow Gaussian distribution $N(\mu _1,\sigma _1^2)$. Similarly the result of vacuum state interference from port ${c_1}$ and $LO_2$ from port ${c_2}$ will follow Gaussian distribution $N(\mu _2,\sigma _2^2)$. So their difference will obey $N(\mu _1-\mu _2,\sigma _1^2+\sigma _2^2)$. As is known, the phase jitter of the two arms is a slow process, so in a short time interval $\tau $, the deviation between the upper and lower arms can be treated as a constant $\mu _1-\mu _2$. Using this data, we can balance the two arms through feedback controlling the phase modulator. Its residual bias caused by the limitation of the feedback control accuracy can be eliminated by an additional subtraction operation.

	The schematic diagram of feedback control is shown in the Figure. \ref{algorithm_pm}. Usually, the phase difference $\Delta \varphi $ between the two arms changes at a speed slower than KHz, which can be compensated to achieve a stable $\Delta \varphi $ when the compensation speed is much faster than the speed of phase jitter. In each compensation period $\tau $, we sum $N$ data sampled during the period and compare the sum value $SUM$ with a desired value. Considering the limited sampling accuracy of the practical $ADC$ and the statistical fluctuation caused by the limited data, we set a decision interval $[A,B]$ to replace the fixed value introduced above. When $SUM$ is in the interval $[A,B]$, the deviation of the output signal is within an acceptable range and an unbiased result can be obtained by subtracting their mean from the $N$ data during $\tau $. 
	
	While in the case that the deviation makes the $SUM$ value out of the interval $[A,B]$, we will adjust the feedback voltage according to the detection results. The output voltage from digital-to-analog converter ($DAC$) convers $2{V_\pi }$ which means 2 times of the half wave voltage of the $PM$ used in the system. When $SUM$ is less than $A$, we reduce the value of $dac\_data$ by $c$ each time, which will result to the reduction of the compensation voltage loaded on $PM$. When the value of $dac\_data$ is less than $c$, we can directly change the value of $dac\_data$ to ${2^n} - c$ due to the two $dac\_data$ value correspond to two close phase modulation result. Conversely, when $SUM$ is larger than the upper bound $B$ of the interval, we increase the value of $dac\_data$ by $c$ each time. When the value of $dac\_data$ is greater than ${2^n} - c$, we set the value of $dac\_data$ as $c$. The above control process makes the value of $SUM$ stable in the interval $[A,B]$, thereby avoiding saturation of the homodyne detector.
	
	\section{\uppercase{EXPERIMENTAL SETUP \& RESULT}}
	
	\noindent We build an all-in-fiber setup with off-the-shelves devices according to the scheme shown in Figure. \ref{system_scheme}. The system includes three main parts: the balance control module, entropy source and an electronic circuit for measurement, calculation and randomness extraction.

	The entropy source consists of a 1550 nm distributed feedback laser (NKT Basic $E15$, line width 100 $Hz$) whose output beam is divided into two beams by $BS_1$ (${\eta _{ab_1}}=3.80dB$, ${\eta _{ab_2}}=3.56dB$). The upper arm is modulated by a phase modulator (EOSPACE, insertion loss ${\eta _{pm}}=3.24dB$, $V_\pi=1.240 V$). The two output signals are coupled into $BS_2$ (${\eta _{c_1d_1}}=3.68dB$, ${\eta _{c_1d_2}}=3.82dB$, ${\eta _{c_2d_1}}=3.76dB$, ${\eta _{c_2d_2}}=3.60dB$). To suppress the deviation of the output signal by the homodyne detector, a feedback control voltage will be loaded on the $PM$. The following $DC$ coupling homodyne detector (Newport, 1817-FC, measurement bandwidth 80 MHz, convertion gain of $PD_1$ $5.55{\rm{ }} \times {10^4} V/W$, convertion gain of $PD_2$ $5.42{\rm{ }} \times {10^4} V/W$) will convert the input optical signal into electrical signal. The $ADC$ card (ADS5463, sampling frequency set as 80 MHz, sampling precision 12 bits and input voltage range 1 VPP) samples the analog signal and quantize it into digital value. The field programmable gate array ($FPGA$, KC705 evaluation board) will sum the sampled $N=1000$ variables and compare the value of sum, $SUM$, with the preset interval [2043000,2053000]. The result of comparison will affect the change in $dac\_data$, which will be converted to the phase compensation voltage through digital-to-analog card ($DAC$, AD9736, sampling precision 14 bits and output voltage range 2.480 VPP) at a speed of 80 $KHz$ when the laser power is set to 5 $mW$. In our experiment, the adjustment step of $dac\_data$ is set as 5.  
	
	The practical discontinuous phase compensation voltage can not meet the requirement of accurate compensation, which results to the compensated signal remain a certain bias. To solve this problem, a subtraction operation between these 1000 variables and their mean will also be implemented on $FPGA$. The result of subtraction is used to randomness estimation and extraction.

	Classical noise introduced by the imperfect devices in the practical system will be controlled by the eavesdropper, Eve, which will result to the information leakage of random numbers, thereby damage the security of the whole system \cite{Bouda2012Weak}. To eliminate the effects of the electrical noise, statistical parameter min-entropy was proposed to quantize the extractable randomness \cite{Ma2012Postprocessing} and a theoretical security proved randomness extractor will be utilized. For the random number generator based on the measurement of the vacuum noise, the outcome of practical measurement $M$ and the noise data $E$ can be obtained when the $LO$ is turned on and turned off separately. $M$ is a combination of the measurement result of quantum noise $Q$ and classical noise $E$. $Q$ and $E$ are assumed to be independent and they both obey Gaussian distribution \cite{Haw2015Maximization}. So the min-entropy of the measurement outcome $M$ conditioned on the existing classical noise $E$ can be given by 
	\begin{equation}
	\label{eq1}
	\begin{split}
	{H_{\min }}\left( {M|E} \right) &=  - {\log _2}\left[ {\mathop {\max }\limits_{e \in E} \mathop {\max }\limits_{m \in M} {P_{M|E}}\left( {m|e} \right)} \right]\\
	&=  - {\log _2}{\left[ {2\pi \left( {\sigma _M^2 - \sigma _E^2} \right)} \right]^{{{ - 1} \mathord{\left/
					{\vphantom {{ - 1} 2}} \right.
					\kern-\nulldelimiterspace} 2}}}\\
	&= {\log _2}{\left( {2\pi \sigma _Q^2} \right)^{{1 \mathord{\left/
					{\vphantom {1 2}} \right.
					\kern-\nulldelimiterspace} 2}}}.
	\end{split}
	\end{equation}
	
	\begin{figure}[t]
		\centering
		\includegraphics[width = 7 cm]{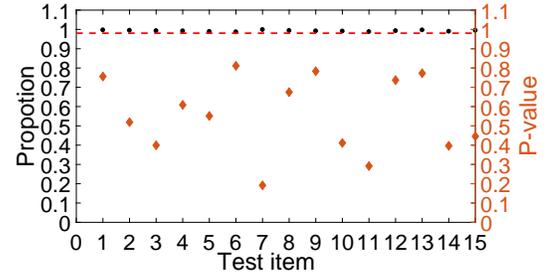}
		\caption{\label{NIST1} Test results of $1000{\rm{ }} \times {10^6}$ random bits using NIST standard statistical test suite. From left to right, the 15 test items shown on the x-axis are named as ‘Frequency’, ‘Block frequency’, ‘Cumulative sums’, ‘Runs’, ‘Longest-run’, ‘Rank’, ‘FFT’, ‘Non-periodic templates’, ‘Overlapping templates’, ‘Universal’, ‘Approximate entropy’, ‘Random excursions’, ‘Random excursions variant’, ‘Serial’ and ‘Linear Complexity’, respectively. On the y-axis, the left and right diagram shows the passing proportion and P-value of each tests, separately. The dotted line shown above is the critical boundary of 0.9805608. }
	\end{figure}
	
	When the $LO$ power is set to 5 $mW$, the measured voltage variance of the raw data ${\sigma _M^2}$ is calculated  as $1.86{\rm{ }} \times {10^5}$. The measured voltage variance of the raw data ${\sigma _E^2}$ is calculated as $166.09$ when the LO power is set to 0 $mW$. Thus the ${H_{\min }}\left( {M|E} \right)$ can be calculated as 10.08 bits per sample or 0.84 bits per raw data bit, which means that $84.0\%$ random bits can be generated from each sample. The final random number output rate will reach 640 Mbps after a real-time randomness extraction based on an improved Toeplitz hashing algorithm proposed in Ref. \cite{ZHENG20186Gbps}. The size of Toeplitz matrix is set as $1920{\rm{ }} \times 2400$ to achieve a security parameter of ${2^{{\rm{ - }}48}}$. Finally, we test their randomness through the $NIST$ standard test suite. The $NIST$ test suite contains 15 statistical tests and each test will output a statistical p-value. The significant level $\alpha $ together with $\beta $ are set as 0.01. $1000{\rm{ }} \times {10^6}$ random bits are used for testing. The sequences will be considered to be random when the proportion of the sequences satisfies p-value $>$ $\beta $ is in the range of $(1 - \beta  - 3{[(1 - \beta )\beta {\rm{/}}N]^{1/2}},1 - \beta  + 3{[(1 - \beta )\beta {\rm{/}}N]^{1/2}})$ \cite{Wang2013_4.5G,Zheng2018Gaussian}.The test results is shown in Figure. \ref{NIST1}.

	\section{CONCLUSIONS}
	
	In this paper, a prototype of bias-free and real-time optical quantum random number generator based on measuring the vacuum fluctuation of quantum state is demonstrated. There are two significant merits of our system favorable for practical applications. First, it can directly reduce the deviation introduced by the unbalanced devices and achieve a bias-free output through compensation and subtraction. Second, the reduction of deviation makes the homodyne detector support a greater LO power to help to achieve a higher min-entropy. Further research can be done by exploring methods to realize accurate compensation and apply the balance technology to other protocols.
	
	\section*{ACKNOWLEDGEMENTS}
	This work was supported in part by the Key Program of National Natural Science Foundation of China under Grants 61531003, the National Natural Science Foundation under Grants 61427813, the National Basic Research Program of China under Grants 2014CB340102, China Postdoctoral Science Foundation under Grant 2018M630116, and the Fund of State Key Laboratory of Information Photonics and Optical Communications.


\end{document}